\documentclass{emulateapj}

\usepackage{color} 
\usepackage{amsmath}
\usepackage{graphicx}
\usepackage{tabularx}
\usepackage{subfigure}
\usepackage{longtable}
\usepackage{multirow}
\usepackage{url}
\usepackage{hyperref}
\usepackage{ulem}

\newcommand\M{\rule{0pt}{2.3ex}}       

\newcommand{\cR}{\mathcal{R}}
\newcommand{\n}{{\rm{N}}}
\newcommand{\bX}{\boldsymbol{X}}
\newcommand{\bY}{\boldsymbol{Y}}
\newcommand{\V}{\mathcal{V}}

\newcommand{\MCz}{{\rm MC}$\_${\rm z}}
\newcommand{\MCztwo}{{\rm MC}$\_${\rm z}{\hskip1pt}2}
\newcommand{\MCzml}{{\rm MC}$\_${\rm z}$\_${\rm ml}}

\shorttitle{Interpreting the Distance Correlation Results for the COMBO-17 Survey}
\shortauthors{Richards, Richards \& Mart\'inez-G\'omez}

\begin{document}

\title{Interpreting the Distance Correlation Results for the COMBO-17 Survey}

\author{Mercedes T. Richards\altaffilmark{1,2}, Donald St. P. Richards\altaffilmark{2,3}, Elizabeth Mart\'inez-G\'omez\altaffilmark{4} } 

\affil{
{$^1$}Department of Astronomy \& Astrophysics, Pennsylvania State University, University Park, PA 16802, USA, mrichards@astro.psu.edu \\
{$^2$}Institut f\"ur Angewandte Mathematik, Ruprecht-Karls-Universit\"at Heidelberg, Im Neuenheimer Feld 294, 69120 Heidelberg, Germany
{$^3$}Department of Statistics, Pennsylvania State University, University Park, PA 16802, USA, richards@stat.psu.edu \\
{$^4$}Department of Statistics, Instituto Tecnol\'ogico Aut\'onomo de M\'exico, Del. \'Alvaro Obreg\'on, 04510, M\'exico D. F., Mexico, elizabeth.martinez@itam.mx \\
}

\begin{abstract}
The accurate classification of galaxies in large-sample astrophysical databases of galaxy clusters depends sensitively on the ability to distinguish between morphological types, especially at higher redshifts.  This capability can be enhanced through a new statistical measure of association and correlation, called the {\it distance correlation coefficient}, which has more statistical power to detect associations than does the classical Pearson measure of linear relationships between two variables.  The distance correlation measure offers a more precise alternative to the classical measure since it is capable of detecting nonlinear relationships that may appear in astrophysical applications.  We showed recently that the comparison between the distance and Pearson correlation coefficients can be used effectively to isolate potential outliers in various galaxy datasets, and this comparison has the ability to confirm the level of accuracy associated with the data.  In this work, we elucidate the advantages of distance correlation when applied to large databases.  We illustrate how the distance correlation measure can be used effectively as a tool to confirm nonlinear relationships between various variables in the COMBO-17 database, including the lengths of the major and minor axes, and the alternative redshift distribution.  For these outlier pairs, the distance correlation coefficient is routinely higher than the Pearson coefficient since it is easier to detect nonlinear relationships with distance correlation.  The V-shaped scatter plots of Pearson versus distance correlation coefficients also reveal the patterns with increasing redshift and the contributions of different galaxy types within each redshift range.
\end{abstract}


\section{Introduction}

The classification of galaxies has been of great interest for decades, and the ability to distinguish between morphological types is pertinent, especially  at higher redshifts.  \citet{kin96} created ultraviolet to near-infrared spectral energy distributions for local galaxies to establish a relationship between morphological type and spectral energy distribution for the mainstream Hubble classification of elliptical, bulge, lenticular, and Sa-Sc spiral galaxies, as well as starburst galaxies.  The overall lack of emission lines hinders the accurate classification of redshifts for more distant galaxies, except in the case of starburst galaxies with optical spectra \citep{kin96}.  In addition, redshift surveys typically measure inadequate luminosity functions for spiral, elliptical, and lenticular galaxies because of contamination by dwarf galaxies \citep{del03}; hence there is a need to more accurately distinguish between the giant and dwarf galaxy types.  These concerns suggest that there remains a need to identify objects that may have been misclassified using traditional methods.

Mart{\'i}nez-G{\'o}mez, Richards \& Richards (2014) recently applied a new statistical measure of association and correlation, called the {\it distance correlation coefficient}, to the COMBO-17 database.  This was the first application of distance correlation to astrophysical data.   Some advantages of the distance correlation measure are that it can detect nonlinear associations that are undetectable by the classical Pearson correlation coefficient, it is applicable to random variables of any dimension, and it is zero if and only if the variables are independent \citep{sze09,due14}.  Moreover, \citet{sze09} and \citet{sim12} have shown that the distance correlation coefficient has higher statistical power to detect associations than the well known Pearson correlation coefficient of \citet{pea1895}.

We selected a sample of 15,352 galaxies, with redshifts $0 \le z < 2$ , from the COMBO-17 database.  The analysis was performed on 33 variables, including  mean redshift, luminosity distance, length of major axis, length of minor axis, position angle, magnitudes, and fluxes at various wavelengths and in different observing runs \citep{mar14}.  The Pearson correlation coefficient and distance correlation coefficient were then compared for the 528 pairs of variables corresponding to the selected 33 variables, and the results were displayed in scatter plots.  These plots have distinctive horseshoe or V-shapes, which occur whenever multi-dimensional data are mapped into two dimensions \citep{mar14,dia08}.   They provide a mechanism by which the differences between potential outliers and the remaining data points can be accentuated.  

\begin{table*}[t]
\caption{Galaxy Types and Redshift Ranges}
\centering
\begin{tabular}{llccccc}
\hline\hline \M
  Galaxy  & Kinney et al. (1996)              & Magnitude Range based on 				     & \multicolumn{4}{c}{Number of Galaxies, N} \M \\ 
 \cline{4-7}\M
Type     & ~~~~~~~Template            & Wolf et al. (2003a)   			          &  ~~$0 \le z < 0.5$~~  &  ~~$0.5 \le z < 1$~~   & ~~$1 \le z < 2$~~ & Total \\[0.5ex]\hline \M
Type 1  & Elliptical, bulge, S0, Sa     & {$B-r > 1.25$}, {$m_{280}-B \geq 1.1$} &  ~~38 & ~~50 & ~~16 & ~~~104\\
Type 2  & Spiral: Sa, Sbc          & {$B-r > 1.25$}, {$m_{280}-B < 1.1$}      &  ~~45 & ~~19 & ~~~4 & ~~~~68 \\
Type 3  & Spiral: Sbc - SB6                 & {$0.95 < B-r \leq 1.25$} 					          & ~328 & ~277 & ~109 & ~~~714\\
Type 4  & Starburst: SB6 - SB1~~  & {$B-r \leq 0.95$} 							          & 3254 & 9284 & 1928 & 14466\\ [0.5ex]\hline \M
Total     & Elliptical - Starburst    &  													    & 3665 & 9630 & 2057 & 15352\\ [0.5ex]
\hline
\end{tabular}
\label{table1}
\end{table*}

With the availability of this new statistical tool, we can learn more about the COMBO-17 dataset through a deeper examination of the distance correlation results, specifically about the separate classes of galaxies in the sample.  In this Letter, we extend the results of \citet{mar14} to illustrate the ways in which distance correlation can be used to explore general patterns in the  database, and to identify pairs of variables that are associated with deviations from nonlinearity.  In Section 2, we review the definition of the distance correlation coefficient, identify a more extensive set of potential outlier pairs, and examine the patterns with galaxy type and redshift.  In Section 3, we provide a summary of results and conclusions.

\section{Application to the COMBO-17 Database}

The widely-used {\it empirical}, or {\it sample} Pearson correlation coefficient is described by the explicit formula 
\begin{equation}
\label{eq:empirical-Pearson}
r = \frac{\sum_{i=1}^\n (x_i-\bar{x})(y_i-\bar{y})}{\sqrt{\sum_{i=1}^\n (x_i-\bar{x})^2} \cdot \sqrt{\sum_{i=1}^\n (y_i-\bar{y})^2}} \, ,
\end{equation}
where $\bar{x} = \n^{-1}\sum_{i=1}^\n x_i$ and $\bar{y} = \n^{-1}\sum_{i=1}^\n y_i$ are the respective sample means for the random sample $\{(x_i,y_i), i=1,\ldots,\n\}$.

The {\it empirical distance correlation} for the observed data $(\bX,\bY)$ is defined as 
\begin{equation}
\label{eq:empirical-dcor}
\cR_\n(\bX,\bY) = \frac{\V_\n(\bX,\bY)}{\sqrt{\V_\n(\bX)} \cdot \sqrt{\V_\n(\bY)}}
\end{equation}
if both $\V_\n(\bX)$ and $\V_\n(\bY)$ are positive; otherwise, $\cR_\n(\bX,\bY)$ is defined to be 0.  
Here, $\V_\n(\bX,\bY)$ is the {\it empirical distance covariance} for the random sample $(\bX,\bY)$, while $\V_\n(\bX)$ and $\V_\n(\bY)$ are the {\it empirical distance variances} for the data $\bX$ and $\bY$, respectively \citep{mar14}.

Additional details of the theory and applications of the distance correlation measure are provided by \citet{sze09}, and Section 5 of that paper provides several applications which demonstrate the superiority of the distance correlation coefficient over the Pearson correlation coefficient in the context of hypothesis testing for relationships between variables.  Moreover, \citet{sim12} found that, even in the linear case, the Pearson coefficient has only slightly higher statistical power than the distance correlation coefficient.  We note that if $\cR_\n = 1$ then there is a linear relationship between the variables \citep[Theorem 3]{sze07}.  

Table \ref{table1} shows the subdivision of the COMBO-17 data into four galaxy types and three redshift ranges according to their $m_{280}-B$ and $B-r$ colors; in a similar way to that defined by \citet{wol03a}.   Here the galaxy types are based on the  \citet{kin96} galaxy classification template for elliptical, bulge, lenticular, spiral, and starburst galaxies.   The starburst galaxies represent 94\% of our COMBO-17 sample, and hence they dominate our sample.  In addition, the early-type galaxies (elliptical, bulge, S0) have the reddest spectra while the starburst galaxies are much bluer \citep{kin96}; as a consequence, the Type 2 and Type 3 galaxy groups of spiral galaxies may be contaminated substantially by starburst galaxies \citep{del03}.
 
\begin{figure*}[t]
\center
\includegraphics[scale=0.33]{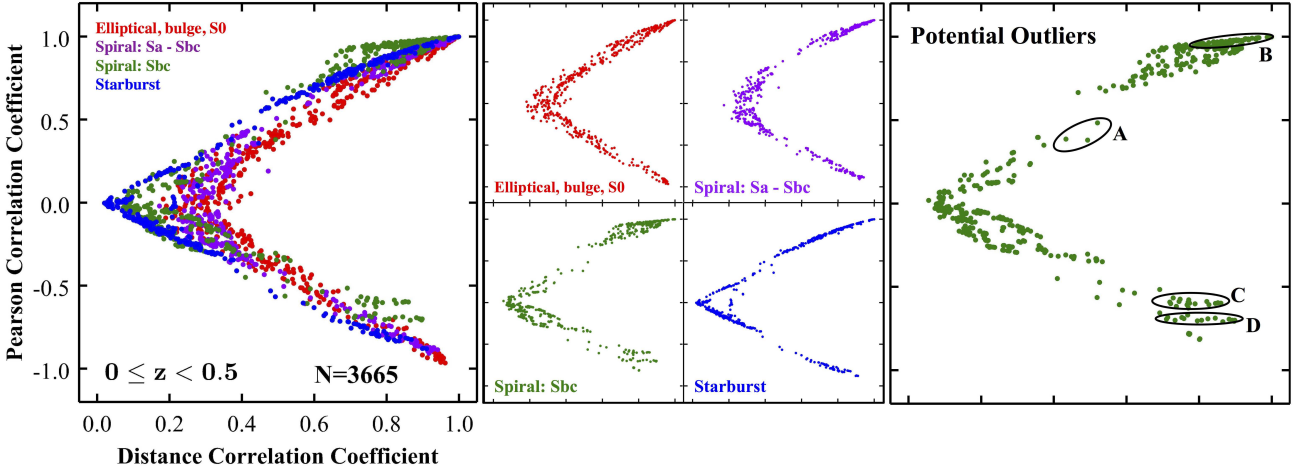}
\caption{Pearson correlation coefficient vs. distance correlation coefficient for the 528 pairs of variables at redshift $0\leq z<0.5$.  The subplots for each galaxy type are shown (four middle frames) along with the superposition of the four subplots (large left frame).   An illustration of locations of potential outlier pairs in the scatter plot for Type 3 spiral Sbc galaxies is also shown (large right frame).  
}
\label{fig1}
\end{figure*}

\begin{figure*}[t!]
\center
\includegraphics[scale=0.33]{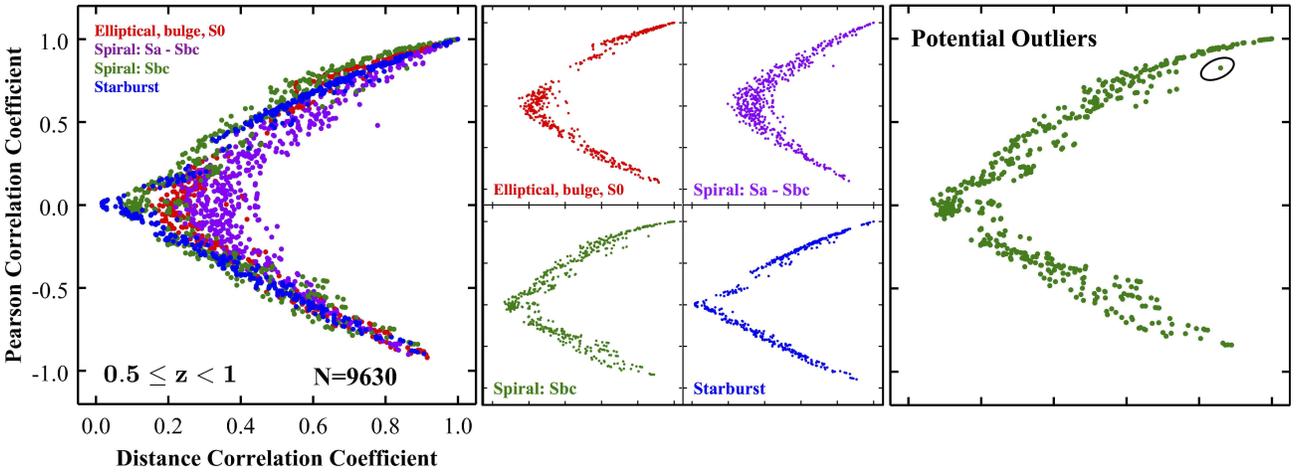}
\caption{Same as Figure \ref{fig1} for redshift $0.5\leq z<1$, and with an expanded view of the plot for Type 3 galaxies (large right frame).
}
\label{fig2}
\end{figure*}

\begin{figure*}[t]
\center
\includegraphics[scale=0.33]{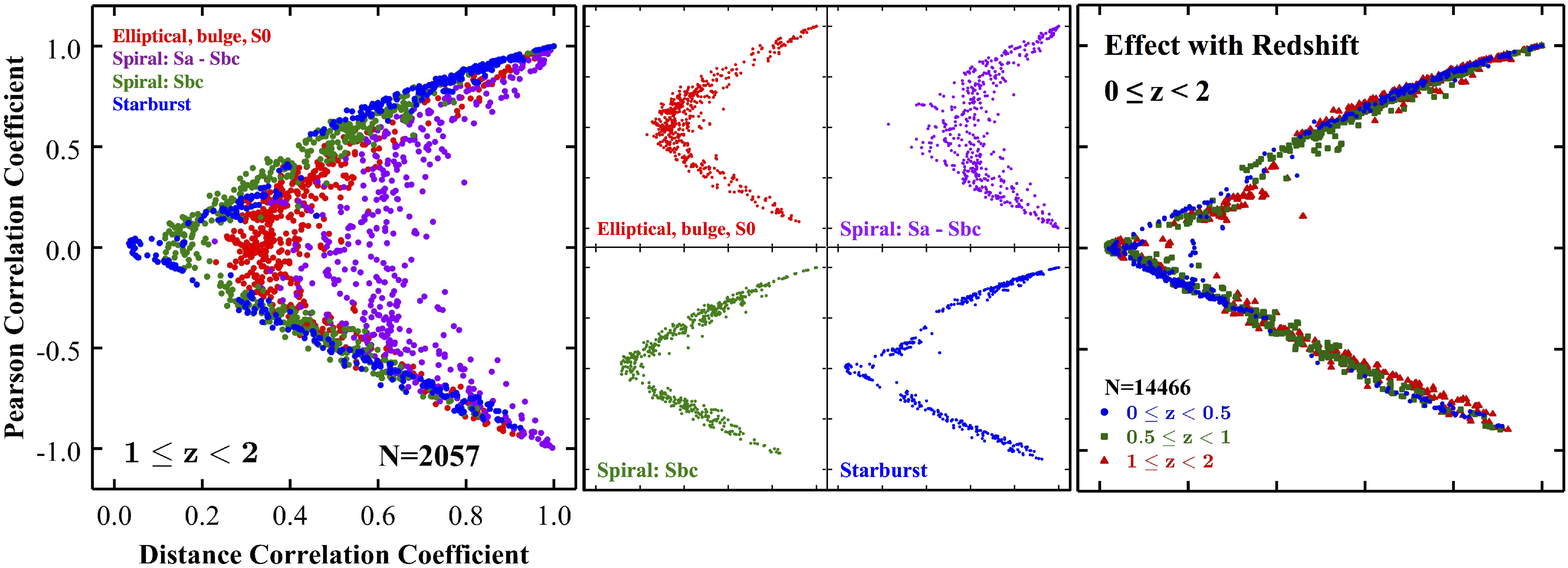}
\caption{Same as left and middle frames of Figure \ref{fig1} for redshift $1\leq z<2$ (left and middle frames).  The effect of redshift on the V-shaped patterns is illustrated in the case of the SB1-SB6 starburst (Type 4) galaxies (large right frame).
}
\label{fig3}
\vskip3pt
\end{figure*}

Figures \ref{fig1} - \ref{fig3} show the V-shaped patterns revealed when the Pearson correlation coefficient, $r$, is compared with the distance correlation coefficient, $\cR_\n$, for the four galaxy types and three redshift ranges.  These figures show different redshift ranges, with four subplots corresponding to galaxy type (four middle frames) and the superposition of these subplots (large left frames).  Also, the V-shaped pattern seen in the left frames would be tighter if all the galaxy types for that redshift range had been combined into a single dataset.  Instead, the left frames in each figure were intended to illustrate the effect of galaxy type on the scatter plot over a fixed redshift range. 

Every point on the graph represents the level of association between the variables in a given pair, and this level of association depends on the number of galaxies, N, in the sample.   The value of N influences the tightness of the V-shaped scatter plot which, in turn, improves our ability to identify potential outlier pairs relative to the general pattern \citep{mar14}.  Moreover, since the number of Type 2 spiral galaxies in each redshift range is small (between 4 and 45), the resulting V-shaped scatter plots for the Sa and Sbc galaxies are fairly diffuse compared to the patterns for the other galaxy types; hence our conclusions about the variables associated with these galaxies is limited.   In contrast, there are thousands of Type 4 starburst galaxies in the sample, with resulting tight V-shaped scatter plots. 

The spread of the points in the V-shaped patterns in  Figures \ref{fig1} -  \ref{fig3} is consistent with the values of N given in Table \ref{table1}. Moreover, they confirm that the galaxy sample is dominated by Type 3 and Type 4 galaxies.

\subsection{Implications of Potential Outliers}

In their introductory application of distance correlation to the COMBO-17 data, \citet{mar14} identified two potential outlier pairs of variables and then realized that these outlier pairs were associated with expected nonlinear relationships.  Hence, the scatter plot of the Pearson and distance correlation measures is an effective tool in isolating potential outlier pairs of variables.  This analysis was extended to identify additional potential outlier pairs and to better understand the results.  The middle frames of Figures \ref{fig1} - \ref{fig3} show that potential outlier pairs of variables can be identified at every redshift.  

Table \ref{table2} provides a list of potential outlier pairs along with the corresponding values of the Pearson correlation coefficient, $r$, and the distance correlation coefficient, $\cR_\n$.
In this table,
MC$\_$z is the mean redshift in distribution $p(z)$;
MC$\_$z2  is the alternative redshift if distribution $p(z)$ is bimodal;
MC$\_$z$\_$ml  is the peak redshift in distribution $p(z)$; 
dl  is the luminosity distance of MC$\_$z;
mu$\_$max is the central surface brightness;
MajAxis is the length of the major axis;
MinAxis is the length of the minor axis;
PA     is the position angle; 
Rmag   is the total $R$-band magnitude;
BjMag is the  absolute magnitude in Johnson $B$;
rsMag is the absolute magnitude in the SDSS $r$-band;
S280Mag is the absolute magnitude in the 280/40 filter;
and  WaF$\_$b  is the photon flux in Filter a in run b  \citep{mar14}.

The first potential outlier pairs identified  were ({\MCztwo},\, {dl}) and ({\MCztwo},\, {\MCz}); {dl} is associated with {\MCz} through Hubble's Law, and  {\MCztwo} is associated with a bimodal probability distribution \citep{mar14}.  Hence, distance correlation had detected the nonlinear nature of the probability distribution.     Table \ref{table2} shows that  the alternative redshift {\MCztwo} is a common variable among the potential outlier pairs at all redshifts and for all galaxy types.  The strongest effect was found for the pair  ({\MCztwo},\, {\MCzml}) for Type 3 Sbc galaxies at $0.5 \le z < 1$ ($r$ =  0.8240, $\cR_\n$ = 0.8590).   

\begin{table}[t]
\caption{Potential Outlier Pairs of Variables }
\hspace{0pt}
\begin{tabularx}{3.35in}{lccrc}
\hline\hline  
Variables                        & Redshift               & Type        & $r$~~  & $\cR_\n$      \M \\ 
[0.5ex] \hline \M
(mu$\_$max,\, MajAxis)	&  $0 \le z < 0.5 $             &  1    &	0.0533 &  0.4245  \\
(mu$\_$max,\, 	rsMag)	&  $0 \le z < 0.5 $             &  1    &   0.0370 &  0.3973 \\
(mu$\_$max,\, 	BjMag)	&  $0 \le z < 0.5 $             &  1    	& 0.0076 &  0.3905   \\
(mu$\_$max,\, 	S280Mag) &  $0 \le z < 0.5 $            &  1    & -0.0320 &  0.3673	 \\
(W518F$\_$E,\,  rsMag)		&  $0 \le z < 0.5 $         &  1     & 	-0.0038 & 0.3634   \\
(W518F$\_$E,\, BjMag)		&  $0 \le z < 0.5 $         &  1        & 0.0020  & 0.3555	\\
(MC$\_$z$\_$ml,\, W462F$\_$E)	&  $0 \le z < 0.5 $    &  1    	&   -0.2964  &  0.3461\\
(MajAxis,\, MinAxis)       &  $0 \le z < 0.5 $             &  2    	& 0.1904  & 	0.4727  \\ 
(MC$\_$z$\_$ml,\, PA)	 &  $0 \le z < 0.5 $             &  2   &		-0.0212   &	0.1719  \\
(rsMag,\,  BjMag) 	           &  $0 \le z < 0.5 $             &  3       & 0.9978 &	0.9987 \\
(MC$\_$z,\, 	dl)					 &  $0 \le z < 0.5 $             &  3       &	0.9994	& 0.9989 \\
(MC$\_$z2,\, mu$\_$max)	 &  $0 \le z < 0.5 $             &  3       &	0.4837  &  0.5200   \\
(MajAxis,\, MinAxis)	            &  $0 \le z < 0.5 $             &  3    	 &   0.3797 &  0.4933 \\
(MC$\_$z2,\,  Rmag)		  &  $0 \le z < 0.5 $             &  3       &	0.3865 & 0.4334    \\
(MC$\_$z,\, MC$\_$z$\_$ml)  &  $0 \le z < 0.5 $             &  4   &	0.0983   &  0.2779 \\
(MC$\_$z2,\, mu$\_$max)   &  $0 \le z < 0.5 $             &  4    &	0.1527   & 0.2417\\
(MC$\_$z2,\,  Rmag) 	&  $0 \le z < 0.5 $             &  4      &	0.1554  & 0.2277  \\
(MC$\_$z2,\,  BF$\_$F)	&  $0 \le z < 0.5 $             &  4    & -0.0231  &  0.2169	   \\
(MC$\_$z2,\,  W571F$\_$E)	&  $0 \le z < 0.5 $             &  4   & -0.1565   &  0.2155 \\
(MC$\_$z2,\,   BF$\_$D)	&  $0 \le z < 0.5 $             &  4      & -0.0240  &  0.2151 \\
(MC$\_$z$\_$ml,\, 	BjMag)	&  $0 \le z < 0.5 $             &  4      & 0.0006  &  0.2135	 \\
(MC$\_$z$\_$ml,\, 	rsMag)	&  $0 \le z < 0.5 $             &  4      & 0.0238 &  0.2123  \\ [0.5ex]
\hline\M
(MinAxis,\, Rmag)	           & $0.5 \le z < 1$   &   1  & -0.0866 & 0.4175	\\
(MinAxis,\, BjMag)		      & $0.5 \le z < 1$   &   1  & 0.0374 &	0.4089	 \\
(MinAxis,\, rsMag)	           & $0.5 \le z < 1$   &   1     & 0.0538 &	0.4031 \\
(MinAxis,\, S280Mag)	      & $0.5 \le z < 1$   &   1   & 0.0035  &	0.3600	 \\
(MinAxis,\, W914F$\_$D)   & $0.5 \le z < 1$   &   1   & 0.1720  &	0.3684	 \\
(MinAxis,\, MC$\_$z2)	      & $0.5 \le z < 1$   &   1    & -0.1859 & 0.3322	\\
(MajAxis,\, Rmag) 	           & $0.5 \le z < 1$   &   1    &	0.0356 & 0.3098   \\
(MajAxis,\, BjMag)	           & $0.5 \le z < 1$   &   1     &	0.0959 &	0.3452 \\
(MajAxis,\, rsMag)	           & $0.5 \le z < 1$   &   1     &  0.1406  & 0.3319 \\
(MC$\_$z2,\, MC$\_$z$\_$ml) & $0.5 \le z < 1$   &   2   & 0.4801  & 0.7781  \\ 
(MC$\_$z2,\, MC$\_$z$\_$ml)   & $0.5 \le z < 1$   &   3 	& 0.8240 &  0.8590 \\
(BjMag,\, S280Mag)	  & $0.5 \le z < 1$   &   3 &	0.6968  &  0.7377 \\
(rsMag,\, S280Mag)  & $0.5 \le z < 1$   &   3 	  & 0.6834  &  0.7191\\
(MC$\_$z2,\, Rmag)	  & $0.5 \le z < 1$   &   3 &	0.6769 &  0.6566  \\
(mu$\_$max	MC$\_$z$\_$ml)  & $0.5 \le z < 1$   &   3  &	   0.6215 & 	0.6536\\
(MajAxis,\, MinAxis)      &   $0.5 \le z < 1$         &  4         & 0.3437  &  0.4221  \\ 
(mu$\_$max,\, MinAxis)    &    $0.5 \le z < 1$   &  4      &  -0.0867  &  0.2238 \\ [0.5ex]
\hline\M
(MC$\_$z2,\, dl)	        & $1 \le z < 2$          &  4      &	0.1566  & 0.4606 \\ 
(MC$\_$z2,\, MC$\_$z)   & $1 \le z < 2$   &  4    &	0.1556  & 0.4592  \\ 
(mu$\_$max,\, MinAxis)	         & $1 \le z < 2$          &  4       &	-0.1381 & 0.2648 \\ [0.5ex]
\hline
\end{tabularx}
\label{table2}
\end{table}

In nearly all instances in Table \ref{table2}, $\cR_\n$ is larger than $|r|$.  It is also noticeable that, in several cases, $|r|$ is close to zero (with $p$-values greater than 5\%) while $\cR_\n$ is greater than zero (with $p$-values less than 5\%).  Examples which display similarly small values of $|r|$ and large values of $\cR_\n$ are described by \citet[subsection 5.2]{sze09} and we conclude, as they did, that the distance correlation coefficient has detected a nonlinear relationship while the Pearson correlation coefficient is unable to detect any relationship.  Therefore, distance correlation is the preferred measure for detecting nonlinear relationships between the potential outlier pairs of variables.

Distance correlation also identified a stronger relationship for the pair MajAxis ($2a$) and MinAxis ($2b$), compared to the Pearson coefficient (Table \ref{table2}); these variables are related to eccentricity ($e$) by the nonlinear quadratic relation:  $b^2 = a^2 (1 - e^2)$.

Figure \ref{fig1} (right frame) illustrates some interesting points and potential outlier regions and  that were examined in greater detail.  This frame focuses on the Type 3 spiral Sbc galaxies over $0.5 \le z < 1$, however the analysis can be applied to any subset of the data.  This scatter plot is distinctive because it resembles the superposition of multiple V-shaped patterns, similar to the left frames in Figures \ref{fig1} - \ref{fig3}, and is likely the result of multiple, and yet distinct, galaxy types within the Type 3 galaxy group.  Since the Type 2 (Sa-Sbc) and Type 3 (Sbc-SB6) groups in the COMBO-17 database may be contaminated with starburst galaxies \citep{del03},  we would need to further refine the classes of spiral galaxies to determine the types corresponding to the different V-shaped patterns seen in the right frames of Figures \ref{fig1} and \ref{fig2}.

Region ``A"  in Figure 1 illustrates the locations of potential outlier pairs, and several were found including the pair (MajAxis,\,MinAxis), which was also identified at other redshifts and for various galaxy types (Table \ref{table2}).

Figure \ref{fig1} (right frame) also shows a nearly horizontal strip of interesting points labeled region ``B" corresponding to pairs with $r \sim 1.0$ and $\cR_\n = 0.9 - 1.0$; these points served as a consistency check on our analysis.  Most of the pairs associated with this strip are composed of B, V, or R magnitudes (in runs D, E, or F)  combined with UV magnitudes at wavelengths 402 - 914 $\mu$m (in runs D or E).  Along the strip, we found 3 pairs with BF$\_$D, 4 pairs with BF$\_$F, 10 pairs with VF$\_$D,  11 pairs with RF$\_$D, 12 pairs with RF$\_$E, and 13 pairs with RF$\_$F.  These associations are a direct consequence of the strong linear relationship between the photon fluxes in different observing runs.  In addition, there are two other pairs along this horizontal strip: ({\MCz},\, dl) with $r = 0.9994$ and $\cR_\n = 0.9989$, which was explained earlier; and  (rsMag,\, BjMag) with $r = 0.9978$ and $\cR_\n$ = 0.9987.

Two more horizontal strips can be seen in the lower part of Figure \ref{fig1} (right frame) corresponding to $r \sim -0.6$  and $r \sim -0.7$ for $\cR_\n \sim 0.8 - 0.9$.
The upper strip with $r \sim -0.6$ (region ``C") has the total R-band magnitude, Rmag, as one variable in the pair, combined with V, R, or UV magnitudes in different runs, and we identified 6 pairs along this strip. The lower strip with $r \sim -0.7$ (region ``D") has the central surface brightness, mu$\_$max, as one variable in the pair, again combined with V, R, or UV magnitudes in different runs, and we found 10 pairs along this strip of potential outliers.  Therefore, these strips are linked to the accuracy to which Rmag and mu$\_$max  are known. 

Some unexpected pairs were also identified as potential outliers, including (rsMag,\, S280Mag), with $\cR_\n$ = 0.7191; and  (BjMag,\, S280Mag) with  $\cR_\n$ =  0.7377.  One possibility is that there are unexpected errors associated with the variable  S280Mag since it appears in both of these outlier pairs.  In addition, these magnitudes are defined over different redshift ranges: BjMag ($z \approx [0.0,1.1]$), rsMag ($z \approx [0.0,0.5]$), and S280Mag ($z \approx [0.25,1.3]$), which might alter the expected relationship between these variables.   Since magnitudes are used to determine distances, the influence of these magnitude definitions over varying redshift ranges should be re-examined.

\subsection{Influence of Redshift}

We used the Type 4 starburst galaxies to illustrate the influence of redshift on the COMBO-17 database because our selected sample of galaxies is dominated by starburst galaxies and there are several thousand of these galaxies in each redshift range.  In addition, potential outlier pairs become more prominent when data sets are combined since larger datasets lead to sharper V-shaped patterns in the scatter plots. 

The superposition of the scatter plots for the starburst galaxies for three redshift ranges is shown in the right frame of Figure \ref{fig3}.   The subplots of the starburst galaxies in Figures \ref{fig1} - \ref{fig3} (middle frames) show a consistent relationship between the redshift range and the sharpness of the V-shaped pattern.   At the lowest redshifts ($0\leq z< 0.5$), the pattern is sharper than at higher redshifts ($0.5\leq z<1$) even though there were fewer galaxies at lower $z$ (N=3254) than at higher $z$ (N=9284).  This sharper distribution at lower $z$ reflects the higher precision in the measurements of variables at lower $z$ relative to those at higher $z$ for starburst galaxies.  

Comparable analyses can be performed for the other galaxy types using similar large samples of galaxies.

\section{Summary and Conclusions}

In this Letter, we extended the work of \citet{mar14} to further explore the application of distance correlation to large galaxy databases, determine the level of association between the measured properties of these galaxies, and examine the consistency of the galaxy classifications across various redshift groups.  

For the application to the COMBO-17 database, we selected a sample of 15,352 galaxies, with redshifts $0 \le z < 2$, and we studied the associations between 528 pairs of variables, based on a selection of 33 variables for each galaxy.  The comparison between the Pearson correlation coefficient and the distance correlation coefficient creates a V-shaped scatter plot, which is an effective tool for identifying potential outlier pairs of variables. 

Most of the potential outliers identified in Table \ref{table2} are directly linked to nonlinear relationships between the variables in the outlier pair.   We also used the V-shaped scatter plots to examine the levels of accuracy associated with three redshift groups of starburst (Type 4) galaxies, and found that the tighter scatter plot for starburst galaxies at lower $z$ confirms the higher precision of these measurements relative to those at higher $z$.

We have shown that distance correlation is more effective than the classical Pearson coefficient in determining the nature of the association between pairs of variables, and we recommend that the distance correlation coefficient be used in place of the Pearson coefficient to determine the relationships between variables.   

Another impact of our results lies in the ability to examine the quality of associations between pairs of variables that are related to large groups of galaxies; this has been achieved through the V-shaped scatter plots that compare the Pearson and distance correlation coefficients, and which depend on the number of galaxies in the sample. If a given galaxy dataset is contaminated by galaxies that do not belong to that group, then the misclassified galaxies may influence the level of association between certain pairs of variables, since some of these relationships may vary with galaxy type.  Hence, knowledge of the level of association between pairs of variables reflects information about the galaxies from which the data were collected, and we can examine the overall quality of the data for groups of galaxies of different types (elliptical, lenticular, spiral, starburst) and also over a range of redshifts.  

In the case of the COMBO-17 database, the Type 2 and Type 3 groups of spiral galaxies are likely to be contaminated substantially by starburst galaxies (de Lapparent 2003); these starburst galaxies also dominate the Type 4 galaxy group in our sample.  Our study revealed further evidence of this contamination by means of the distribution of points in the V-shaped scatter plots for the Type 2 and Type 3 galaxies, and this distribution resembles a superposition of multiple V-shaped patterns (Figures 1 and 2; large right frames).  Similar superpositions of V-shaped patterns were also revealed when we separated the database into the four galaxy types and then plotted the information onto a single graph (Figures \ref{fig1} - \ref{fig3}; large left frames).   Given the increasingly large sizes of galaxy databases, our technique can be used to quickly identify potentially contaminated galaxy groups and then a more detailed study can be performed to identify the specific misclassified galaxies.

\acknowledgements
We thank the referee for helpful comments on the manuscript. This research was partially supported by USA National Science Foundation grants AST-0908440 and DMS-1309808; and a Romberg Guest Professorship at the Heidelberg Graduate School for Mathematical and Computational Methods in the Sciences, funded by German Universities Excellence Initiative grant GSC 220/2.

\end{document}